# How are new citation-based journal indicators adding to the bibliometric toolbox?

Loet Leydesdorff

Amsterdam School of Communications Research (ASCoR), University of Amsterdam, Kloveniersburgwal 48, 1012 CX Amsterdam, The Netherlands;

loet@leydesdorff.net ; http://www.leydesdorff.net

**Abstract**

The launching of Scopus and Google Scholar, and methodological developments in Social Network Analysis have made many more indicators for evaluating journals available than the traditional Impact Factor, Cited Half-life, and Immediacy Index of the ISI. In this study, these new indicators are compared with one another and with the older ones. Do the various indicators measure new dimensions of the citation networks, or are they highly correlated among them? Are they robust and relatively stable over time? Two main dimensions are distinguished—size and impact—which together shape influence. The H-index combines the two dimensions and can also be considered as an indicator of reach (like Indegree). PageRank is mainly an indicator of size, but has important interactions with centrality measures. The Scimago Journal Ranking (SJR) indicator provides an alternative to the Journal Impact Factor, but the computation is less easy.

**Introduction**

In a seminal article about citation analysis as a tool in journal evaluation, Garfield (1972, at p. 476; Garfield & Sher, 1963) advocated the Journal Impact Factor in order to normalize for the expected relation between size and citation frequency. On the basis of a sample of the *Science Citation Index* 1969, he concluded that from 21 to 25 percent of all references cite articles that are less than three old (Martyn & Gilchrist, 1968), and therefore one might define the Journal Impact Factor as the average number of citations in a given year to citable items in the two preceding years. As is well known, this has become the Journal Impact Factor in use by the ISI (Thomson Reuters) and in a large number of evaluation studies.

In later studies, the ISI (Garfield, 1990, 1998) experimented with time windows of five and even fifteen years. The Impact Factor was further formalized and generalized by Frandsen & Rousseau (2005) and by Nicolaisen & Frandsen (2008). However, in a Letter to the Editor of *Information Processing and Management,* Garfield (1986) argued on substantive grounds against the use of five-year Impact Factors: one should not confuse impact with influence (Bensman, 2007). The Impact Factor does not measure the impact or influence of a journal, but of an average item published in that journal (Harter & Nisonger, 1997). In other words, it could imply an "ecological fallacy" to infer from the average quality of trees to the quality of the woods as a whole (Robertson, 1950; Kreft & De Leeuw, 1988).

In the case of journals, the size of the journal also plays a role, or as Garfield adds: "Surely it should be obvious that influence is a combination of impact and productivity." (*Ibid.*, p. 445). Unlike the impact of the average, productivity can be indicated by the total number of documents and/or the total number of citations, publications, etc. Garfield (1979: 149) added that the number of times a journal cites articles it published, or is cited by these articles, provides yet another indicator ("self-citations").

In addition to a breakdown of the citations by year (for the last ten years), the *Journal Citations Reports* of the Institute of Scientific Information (Thomson-Reuters ISI) provide a number of other indices. Most relevant in this context—since citation-based—are the Immediacy Index and the journal's Cited Half-life. The Immediacy Index provides the number of citations an item obtains in the year of publication itself. The cited half-life of a journal is the median age of its articles cited in the current *JCR* year. In other words, half of the citations to the journal are to articles published within the cited half-life.

In a validation study of these indicators against usage data, Bensman (1996; Bensman & Wilder, 1998) concluded that the total number of citations correlates much better with the perceived importance of a journal than with its impact as defined by the *JCR*. In the latter case, the correlation with (LSU) Faculty Rating and (UI) Library Use was 0.36 and 0.37, respectively, while correlations of Total Cites with these usage data ranged between 0.67 and 0.82. He proposed to use "Total Cites" as an important indicator for journal evaluation because "size matters" in human perception. In his sample of 129 chemistry journals, the correlation between Total Cites and Impact Factors was significant, but only 0.43. Leydesdorff (2007a: 28) used Bensman's data to test whether the various indicators were independent using factor analysis, and found two factors (which explained 82% of the variance) when using the various indicators provided by the ISI. The first factor is determined by size, and the second by impact. Faculty scores and usage data correlated with size in this 1993 dataset. Yue *et al*. (2004) found a high correlation between the Impact Factor and the Immediacy Index—as expected because both these indicators refer to the current research front and are normalized by dividing the number of citations by the number of publications. (However, the sets of publications and citations are differently defined for the two indicators.)

More recently, new indicators have been proposed, such as the H-index (Hirsch, 2005) and the so-called Scimago Journal Rank (SJR) using data from Scopus. The H-index is most popular and has been included in the online version of the *Science Citation Index-Epanded* of the ISI for any set of documents. The H-index was originally defined at the author level: a scholar with an index of *h* has published *h* papers each of which has been cited by others at least *h* times. However, like the other measures it can be applied to any document set (Braun *et al*., 2006; Van Raan, 2006). Unlike the other measures, the H-index is time-dependent, or one might say dynamic.[1] Given the advent of Internet-based databases such as Google Scholar, this continuous update could also be considered as an advantage. Its proponents claim that the H-index reflects both the number of publications ("productivity") and the number of citations per publication ("impact"). Ever since its introduction a number of derived indicators have been proposed, like the G-index, the AR-index, etc., which improve on some of the shortcomings of the H-index (e.g., Egghe & Rao, 2008; Rousseau, 2008).

---

[1] The H-index is not necessarily dynamic, but it is most often used in this way. Like the impact factor, the H-index provides a framework for the evaluation of a document set.

In 2004, Elsevier launched the Scopus database as an alternative to the ISI databases. Scopus covers more journals than the *Science Citation Index*, the *Social Science Citation Index,* and the *Arts & Humanities Citation Index* of the ISI combined. In both databases, however, the inclusion of journals is based on both quantitative information about journals and qualitative expert judgements (Garfield, 1990; Testa, 1997). Criteria are not externally transparent, but this seems legitimate because of the commercial interests at stake for journal publishing houses. For example, during the early 1980s the ISI resisted pressure from Unesco to include more journals from lesser developed countries (Moravčik, 1984, 1985; Gaillard, 1992; Maricic, 1997). Scopus includes many more Chinese journals than the ISI database, but more recently there seems to be an agreement to include more regional (including Chinese) journals in the ISI domain (e.g., http://globalhighered.wordpress.com/2008/05/29/thomson-scientific-china/ ).

The two databases (*Scopus* and the *Science Citation Indices*) are both overlapping and complementary (Meho & Yang, 2006; Visser & Moed, 2008). The third database, of course, is Google Scholar, which was also launched in 2004. Google Scholar is based on crawling the Internet for scientific literature, and inclusion criteria are relaxed (albeit also not transparent): authors and publishers of scientific articles are encouraged to submit their materials. Given the design of this database which returns hits along a decreasing order of citations, the results almost invite the user to consider the H-index for a set by paging down the list till one reaches the point where the number of citations breaks even with the sequence number. The H-index and Google Scholar are thus most apt to relate to each other.

The search engine Google itself uses PageRank as an algorithm for sorting pages when displaying the search results (Page *et al*., 1998; Brin & Page, 1998). PageRank is derived from the Influence Weights that Pinski & Narin (1976) originally proposed as an indicator of journal status (Garfield, 1979). PageRank is included in the Korean software package NetMiner; the original program of Brin & Page is freely available in the Network Workbench Tool at http://nwb.slis.indiana.edu/ (NWB Team, 2006). I shall include PageRank in the comparison among journal indicators below.

Based on the Scopus database, the Scimago research group of the Universities of Granada, Extremadura and Carlos III in Madrid (http://www.atlasofscience.net) has developed the so-called Scimago Journal and Country Rank system at http://www.scimagojr.com/index.php. Particularly relevant for my research question is the set of journal indicators made available at http://www.scimagojr.com/journalrank.php . All data is brought online and available for further research. The Scimago Journal Rank (SJR) can be considered as an equivalent in the Scopus domain to the Journal Impact Factor in the ISI domain (Falagas *et al*., 2008).

My first research question is: do these indicators measure a common dimension in the data? It would have been nice to include usage data as currently collected in the so-called *mesur project* of the Los Alamos National Laboratory in this comparison, but unfortunately this data is hitherto not available for further research (Bollen & Van de Sompel, 2006; Bollen *et al*., 2008; Bollen, *personal communication*, April 25, 2008).

Are the various indicators complementary or essentially measuring the same underlying dimension? Secondly, how do they relate to other measures of journal impact, influence, size, etc.? Thirdly, how stable are these measures over time? If considerable error were introduced by an algorithm, one would expect the resulting measures to be less stable than the raw data, e.g., Total Cites. For this reason, I also compare the data for 2007 with similar data for 2006 and provide the auto-correlations between data for these two years. In addition to comparing the journal indicators among them, I extend the analysis with some network indicators from social network analysis such as centrality measures using the same datasets.

**Methods and materials**

The data of the *Journal Citations Reports* of the *Science Citation Index* and the *Social Science Citation Index* are available online at the ISI Web of Knowledge (http://www.isiknowledge.com). In this project the data was harvested from the CD-Rom version of the databases, which are otherwise similar to the electronic versions but easier to manipulate using relational database management. The two datasets (for the *Science*

*Citation Index* and the *Social Science Citation Index*, respectively) are combined so that one can correct for the overlap between them (342 journals in 2007 and 321 in 2006, respectively). The datasets contain 7,940 journals in 2007 and 7,611 in 2006.

With dedicated software a full citation matrix for these datasets can be constructed. For example, in 2007 one can obtain a 7940 x 7940 journal matrix with the *cited* journals on one axis and the same journals *citing* on the other. This matrix represents a valued and directed graph. The matrix can be stored, for example, as an SPSS systems file. However, of the 7,940 x 7,940 = 63,043,600 possible cells, only 1,460,847 (2.32%) are not empty. Using a list format (like edgelist in UCINet or Arcs in Pajek) one can hence store this data more efficiently. Using Pajek (or UCINet) one can compute centrality measures like degree centrality, betweenness centrality, and closeness centrality, both in the cited and the citing dimensions of these asymmetrical matrices. As noted, one can also provide the PageRanks of the journals. I have made these various indicators available at http://www.leydesdorff.net/jcr07/centrality/index.htm.

The indicators based on data from the Scopus database have been made conveniently available by the Scimago team at the websites for the respective years as MS-Excel files. The data contains the total numbers of documents, references, and citations (with a breakdown for the last three years), the SJR value, and the H-index for all journals in the set. However, Total Cites are provided only for the last three years. The help-file formulates: "**Total Cites (3years)/Total Cites**: Total of document citations received by a journal in a 3 year period. This indicator is estimated taking into account of all types of documents contained in a journal in the selected year and the bibliographical references they include to any document published in the three previous years." In another context, this indicator is compared with the Impact Factor Numerator based on using a two-year citation window retrospectively (Bensman & Leydesdorff, in preparation). I use this indicator as the best proxy for "Total Cites" available in the Scopus database.

Another problem with the Scimago/Scopus database is the H-index provided for each of the years. Using different dates for the download, I noted that these H-indices are updated, presumably quarterly, and then retrospectively also in the previous years. At least, we found the same values for every year at one moment in time (October 1, 2008), and different values between searches in April 2008 (when downloading the data for 2006) and June 2008 (data for 2007). However, the distribution of H-indices should vary between the years, for example, because of variations in the database coverage. In 2007, the Scopus database covered 13,686 journals as against 13,210 in 2006.

More worrisome than this understandable updating of the H-values from the perspective of hindsight—because this dynamic increase is a well-known problem of the H-index (Jin *et al*., 2007)—is the apparent, but unexplained updating of the SJR values over time. The *Annual Review of Immunology*, for example, which ranks highest on this indicator in both 2006 and 2007, had an SJR 2006 of 23.740 on October 1, 2008, but only 22.439 in April 2008. Similarly, the second journal listed on this ranking, the *Annual Review of Biochemistry*, had an SJR 2006 of 16.796 in September but 16.100 in April. The SJR values in 2007 are lower for both these journals, namely 20.893 and 15.691, respectively.

Might these values also increase with time? The formula for calculating the SJR-values is provided at http://www.scimagojr.com/SCImagoJournalRank.pdf (accessed on October 4, 2008), but it does not suggest a dynamic perspective.

In this study, I used the values for 2006 as they were downloaded on April 23, 2008, and the values for 2007 as downloaded on June 24, 2008, since I assumed that the databases are reloaded on the occasion of the yearly update (early June). (The help-file mentions updating periodically.) Anyhow, for the statistics—which are the subject of this study— these relatively small differences are probably not so important.

In the second part of the study, I also use network centrality measures (Freeman, 1978/1979; Hanneman & Riddle, 2005; Leydesdorff, 2007b). Six possible centrality measures (degree, betweenness, closeness, in both the "cited" and "citing" dimensions) can be calculated using Pajek. PageRank is not included in Pajek, but it is in several other programs. Both NetMiner and the original program of Brin & Page (as included in the Network Workbench at http://nwb.slis.indiana.edu) were used to calculate PageRanks for the journals in 2006 and 2007. The results were identical for the (default) damping factor $d = 0.15$ and ten bins. I also used other parameters; this will be discussed in the results section.

Let me first focus on the static analysis for 2007. After a factor analysis and correlation analysis of the various journal indicators for 2007, I shall add the centrality measures to the domain in order to see whether and how they contribute to understanding the matrix and the indicators. The same analysis was also done for 2006. The results are virtually similar which suggests a high degree of stability in the structure of this set of indicators. Thereafter, the various indicators are auto-correlated between the corresponding values for 2006 and 2007.

Because significance testing is dependent on the number of cases, and the number of cases in the Scopus database is much larger than in the combined ISI-databases (the *Science Citation Index* and the *Social Science Citation Index*), I used the overlap between the two databases (Table 1). The databases were matched using their full titles.

| Number of journals | ISI databases | Scopus | Overlap |
|---|---|---|---|
| 2006 | 7,611 | 13,210 | 6,045 |
| 2007 | 7,940 | 13,686 | 6,210 |

**Table 1**: number of journals in the ISI-databases combined, Scopus, and the overlap.

The dynamic analysis (in terms of auto-correlations for different years) is based on the 5,861 journals included in both databases in both 2006 and 2007.

**Results**

*Journal indicators*

Let us first limit the analysis to the typical journal indicators. These are the size indicators such as total numbers of citable documents, total numbers of references and citations, the Impact Factor, Immediacy Index, Cited Half-life, the SJR, and the H-index. I also added

PageRank to this set since this indicator is commonly known because of its use in the Google database. Furthermore, PageRank can be considered as an inherited revival of the so-called Influence Weights proposed in the 1970s by Pinsky and Narin (1976). However, PageRank is dependent on parameter choices like the damping factor $d$. While $d = 0.15$ is the default value, some authors recommend $d = 0.85$. Ma *et al.* (2008, at p. 803) argued for $d = 0.5$ in the case of citation analysis (as different from hyperlink analysis). Thus, I first tested the correlation in the ranking using different values of the parameter: I found strong correlations (Pearson's $r = 0.942$, and Spearman's $\rho = 0.934$) when comparing the two extremes of $d = 0.15$ and $d = 0.85$ ($p < 0.01$; $N = 6,160$). In the factor analysis, it made *no* difference which damping factor was used for the PageRank analysis. Although PageRank is a network indicator, I decided to include it as a journal indicator in the first round of the analysis because of its origins in the journal domain (Pinski & Narin, 1976).[2]

A three-factor solution explains 82.8% of the common variance. (These three components explain 55.1, 19.5, and 8.2%, respectively.) Table 2 provides the factor solution based on Varimax rotation and Kaiser normalization. Figure 1 shows the scatterplot using the two main components. Obviously, "Cited Half-life" provides the third (latent) dimension in this data-structure. This variable does load (slightly) negatively on the first two dimensions.

---

[2] Bollen & Van de Sompel (2006) used a weighted PageRank on the aggregated journal-journal citation matrix of the ISI; Bergstrom (2007) used the eigenfactor-value as a proxy for PageRank and posted the results online at http://www.eigenfactor.com .

**Rotated Component Matrix(a)**

|  | Component | | |
|---|---|---|---|
|  | 1 | 2 | 3 |
| Total Citing (ISI) | .951 | .147 |  |
| Total Docs. (ISI) | .944 |  |  |
| Total Citing (Scopus) | .942 | .152 |  |
| Total Docs. (Scopus) | .932 | .125 |  |
| Self-Citations (ISI) | .831 |  |  |
| Total Cited (Scopus) | .773 | .456 | .139 |
| Total Cited (ISI) | .763 | .407 | .213 |
| PageRank ($d = 0.85$) | .729 |  | -.112 |
| Impact Factor | .135 | .951 |  |
| Immediacy Index |  | .895 |  |
| SJR |  | .878 |  |
| H-index | .577 | .671 | .222 |
| Cited Half-life |  |  | .963 |

Extraction Method: Principal Component Analysis.
Rotation Method: Varimax with Kaiser Normalization.
a  Rotation converged in 4 iterations.

**Table 2**: Results of the factor analysis for journal indicators.

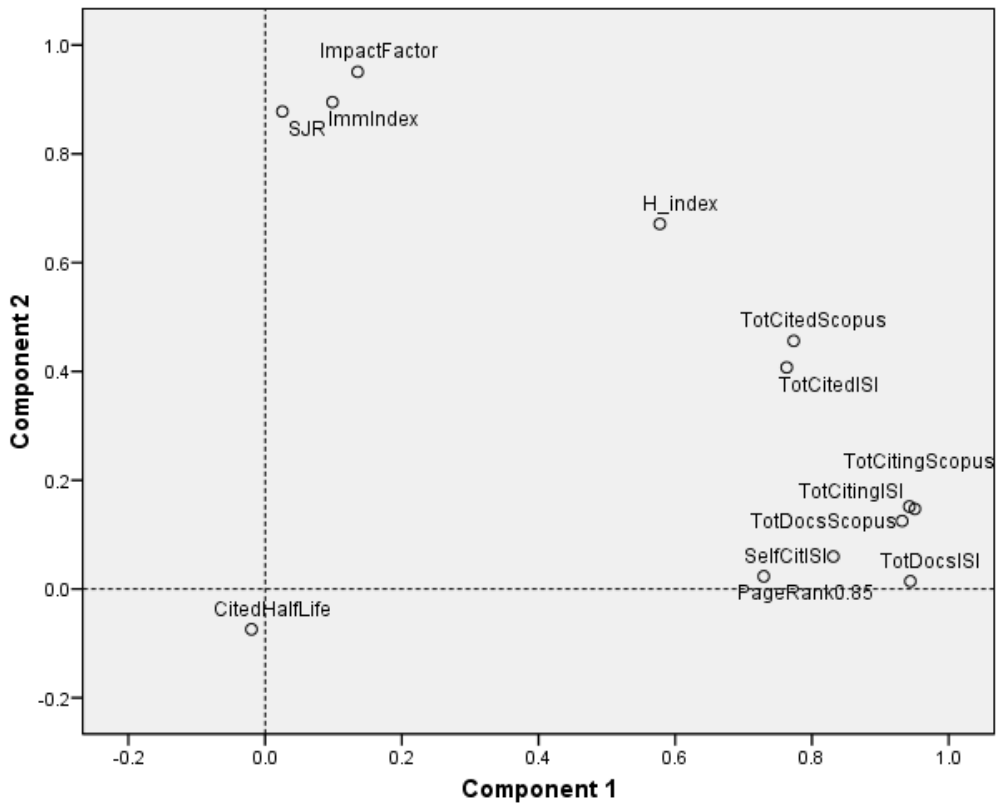

**Figure 1**: Scatterplot of the journal indicators on the two main components; $N = 6,210$.

Component One represents the size factor, Component Two the impact. We can follow Garfield (1986) by concluding that influence is indeed a combination of productivity (size) and impact. Among this set of indicators, PageRank Centrality is obviously a size indicator. The H-index, however, combines the two dimensions more than any of the other indicators. This characteristic may explain its almost instant popularity among research evaluators.[3]

Table 2 further informs us that the numbers of references and the numbers of documents are highly correlated, and these values in either database can be used as indicators of the size dimension. Self-citations also follow on this size dimension. The numbers of citations ("Total Cited") in the Scopus and the ISI-database are correlated with $r = 0.94$ ($p < 0.01$; $N = 6,160$) despite the different definition of Total Cites in the Scopus database. By limiting the citation window to the last three years, Total Cites in the Scopus database loads a bit higher on the impact dimension because of the increase in focus on the research front.

Another way of designating the two factors would be to consider the first size-related factor as indicating the archival function, and the second one as indicating the research front (Price, 1965). As mentioned, the Impact Factor and the Immediacy Index correlate closely, but the new ranking indicator SJR also correlates in this dimension. These correlations are provided in Table 3. (The significance at the 1% level of all correlations is a consequence of the large numbers; $N > 6,000$.)

---

[3] A four-factor solution—the fourth factor explains an additional 5.3% of the common variance—further differentiates between the Total Cited, on the one hand, and the Total Citing and the PageRank, on the other.

**Correlations**

|  |  | Impact Factor | Immediacy Index | SJR |
|---|---|---|---|---|
| Impact Factor | Pearson Correlation | 1 | .877(**) | .796(**) |
|  | Sig. (2-tailed) |  | .000 | .000 |
|  | N | 6158 | 6102 | 6158 |
| Immediacy Index | Pearson Correlation | .877(**) | 1 | .671(**) |
|  | Sig. (2-tailed) | .000 |  | .000 |
|  | N | 6102 | 6104 | 6104 |
| SJR | Pearson Correlation | .796(**) | .671(**) | 1 |
|  | Sig. (2-tailed) | .000 | .000 |  |
|  | N | 6158 | 6104 | 6160 |

** Correlation is significant at the 0.01 level (2-tailed).

**Table 3**: Correlations among the Impact Factor, Immediacy Index, and SJR in the overlap between the journals included in the ISI database and Scopus.

The Impact Factor correlates less with the SJR measure than with the Immediacy Index; as Figure 1 shows, SJR is slightly more orthogonal to the size dimension than the Impact Factor. This means that it normalizes for size a bit more strongly than the Impact Factor already does.

Recently, Zitt & Small (2008) proposed the Audience Factor (AF) as an alternative to the Impact Factor (IF). These authors used 5,284 journals from the *Journal Citations Reports* 2006 of the *Science Citation Index*, of which 4,277 belong to the sets studied in this research. Furthermore, they used five-year citation windows (AF5 and IF5, respectively). Using these 4,277 journals and including these two variables, I generated Figure 2, which is based on adding these two indicators (AF5 and IF5) to the set.

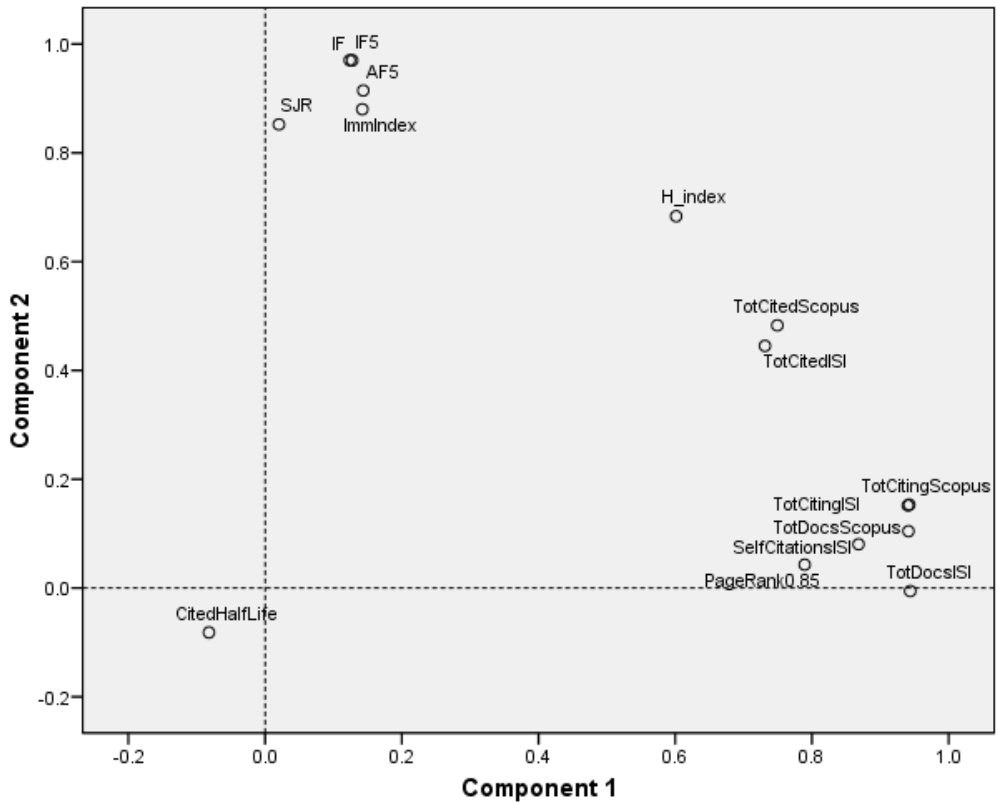

**Figure 2**: Scatterplot of the journal indicators 2006 on the two main components; audience factor added; $N = 4,277$.

|  |  | IF5 | IF | ImmIndex | SJR |
|---|---|---|---|---|---|
| AF5 | Pearson Correlation | .956(**) | .921(**) | .790(**) | .685(**) |
|  | Sig. (2-tailed) | .000 | .000 | .000 | .000 |
|  | N | 4277 | 4277 | 4191 | 4277 |

** Correlation is significant at the 0.01 level (2-tailed).

**Table 4**: Pearson correlation coefficients for the Audience Factor (AF5) with the other impact indicators.

Table 4 provides the correlations of the AF5 with the other impact indicators. A comparison of Figure 2 with Figure 1 shows (1) how robust the two factor solutions are when comparing 2006 and 2007, and (2) that the Audience Factor does not add structurally to the set of indicators already available.

*Network indicators*

Let me now add the six centrality indicators that were derived from social network analysis to this system of journal indicators. A three-factor solution in this case explains 74.5% of the common variance, but the screeplot indicates that a five-factor solution should be considered (85.7%; eigenvalues > 1.0). If we choose this five-factor solution, the network indicators add a third and a fourth dimension to the three-factor solution of the journal indicators discussed above. The "Cited Half-life" remains a final (fifth) factor (to be discussed below).

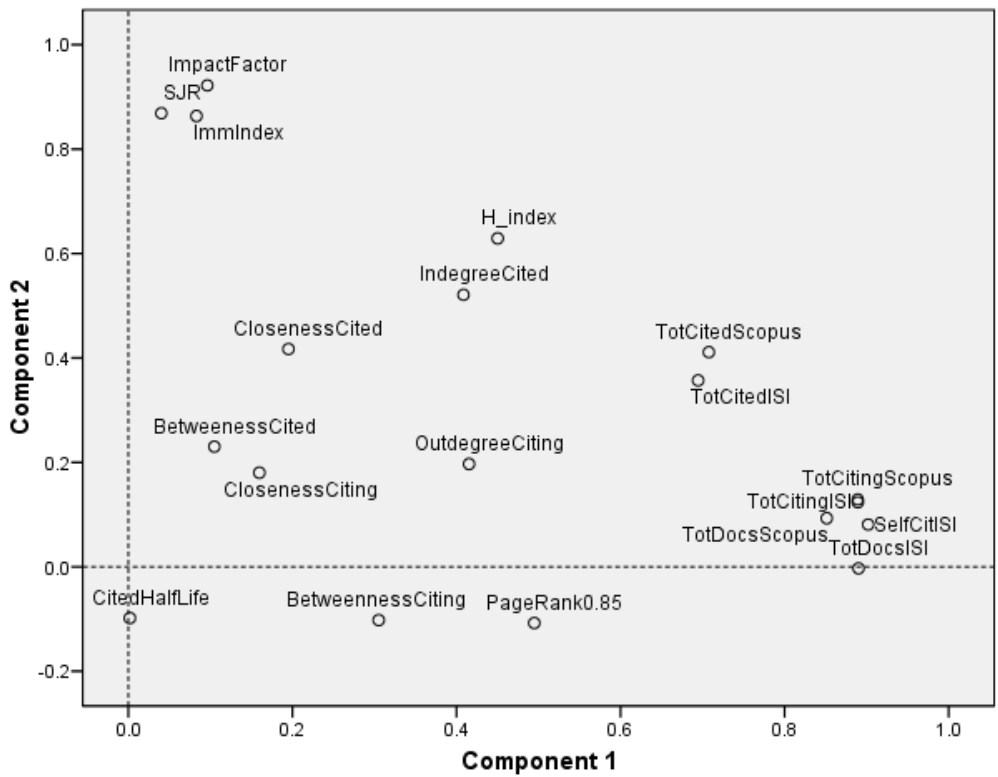

**Figure 3:** Scatterplot of the two main components of a five-factor solution including the network indicators ($N = 6,210$).

Figure 3 illustrates how the network indicators can be projected using the two axes spanned by the journal indicators as the first two components. The only journal indicator that is sensitive to the addition of the network indicators is PageRank, because this is an indicator based on graph theory. Because of the formula of the SJR—provided at http://www.scimagojr.com/SCImagoJournalRank.pdf (accessed on October 4, 2008)—one would expect this indicator also to be sensitive to the addition of network indicators, but this is hardly the case. In other words, the SJR is indeed a journal impact indicator.

**Rotated Component Matrix(a)**

|  | Component |  |  |  |  |
|---|---|---|---|---|---|
|  | 1 | 2 | 3 | 4 | 5 |
| Self-Citations (ISI) | .902 |  |  |  |  |
| Total Docs. (ISI) | .890 |  | .323 |  |  |
| Total Citing (ISI) | .889 | .124 | .366 |  |  |
| Total Citing (Scopus) | .889 | .129 | .338 |  |  |
| Total Docs. (Scopus) | .851 |  | .324 | .194 |  |
| Total Cited (Scopus) | .708 | .411 | .128 | .459 |  |
| Total Cited (ISI) | .695 | .357 |  | .540 | .128 |
| Impact Factor |  | .922 | .185 | .102 |  |
| SJR |  | .869 |  |  |  |
| Immediacy Index |  | .863 |  | .130 |  |
| H-index | .450 | .629 | .375 | .300 | .246 |
| Indegree Cited | .408 | .521 | .520 | .344 | .220 |
| Closeness Citing | .160 | .180 | .840 |  |  |
| Outdegree Citing | .415 | .197 | .800 | .123 |  |
| Closeness Cited | .195 | .417 | .762 |  | .182 |
| PageRank ($d = 0.85$) | .495 | -.108 | .636 | .294 | -.112 |
| Betweenness Citing | .305 | -.102 | .554 | .550 | -.119 |
| Betweenness Cited | .105 | .230 |  | .905 |  |
| Cited Half-life |  |  |  |  | .947 |

Extraction Method: Principal Component Analysis.
Rotation Method: Varimax with Kaiser Normalization.
a Rotation converged in 6 iterations.

**Table 5**: Five-factor solution for the set of journal and network indicators ($N = 6{,}210$).

The patterns among the network indicators are complex (Table 5). First, it is to be noticed that Indegree (that is, the number of links in the cited dimension) correlates highly with the H-index ($r = 0.89$). The Indegree can also be considered as a measure of reach, that is, the number of citation relations but without weighting for the number of citations on these relations. As could be expected, betweenness centrality is different from closeness and degree centrality (Freeman, 1977; Leydesdorff, 2007b). PageRank shares interfactorial complexity with the H-index and Indegree, but unlike the latter two PageRank correlates negatively with the indicators on the impact dimension (Factor 2).

| *Pearson correlations* | Indegree Cited | Outdegree Citing | Betweenness Cited | Betweenness Citing | Closeness Cited | Closeness Citing |
|---|---|---|---|---|---|---|
| **Impact Indicators** | | | | | | |
| Impact Factor | .602 | .399 | .289 | .194 | .448 | .289 |
| Immediacy Index | .507 | .298 | .282 | .162 | .347 | .219 |
| *Cited Half-life* | *.093* | *-.130* | *.048* | *-.005* | *.012* | *-.167* |
| **New Indicators** | | | | | | |
| SJR | .464 | .253 | .186 | .077 | .295 | .176 |
| H-index | .890 | .635 | .460 | .403 | .632 | .414 |
| PageRank ($d = 0.85$) | .547 | .744 | .224 | .772 | .357 | .430 |
| **Size indicators** | | | | | | |
| Total Docs. (ISI) | .549 | .629 | .182 | .498 | .375 | .396 |
| Total Docs. (Scopus) | .626 | .647 | .297 | .516 | .409 | .381 |
| Total Cited (ISI) | .726 | .491 | .643 | .437 | .359 | .244 |
| Total Cited (Scopus) | .736 | .545 | .553 | .428 | .379 | .275 |
| Total Citing (ISI) | .645 | .724 | .206 | .534 | .433 | .405 |
| Total Citing (Scopus) | .623 | .689 | .204 | .517 | .419 | .384 |
| Self-Citations (ISI) | .409 | .385 | .160 | .338 | .253 | .212 |

**Table 6**: Pearson correlations among the network and journal indicators. (All correlations are significant at the level of $p < 0.01$.)

Table 6 provides an overview of the correlations between the two types of indicators, namely the network indicators based on graph theory, and the traditional and new journal indicators. More than the other journal indicators, "Citation Half-life" does *not* correlate (positively or negatively) with the network indicators because of its focus on the time dimension. Citations are conceptually related to time, and journal indicators take this into account by using for example a two-year citation window (as in the case of the Impact Factor). Network indicators, however, do not take time into account, but are momentary. The historical dimension can be programmed into the network approach, but this has to be done explicitly (Pudovkin & Garfield, 2002; Lucio-Arias & Leydesdorff, 2008; Leydesdorff & Schank, 2008).

*Stability and change*

Table 7 provides the auto-correlations for the various indicators for the years 2006 and 2007, using the 5,861 journals included in both databases in both years.

|  | Pearson's r | Spearman's ρ |
|---|---|---|
| ***ISI-Journal Indicators*** | | |
| Impact Factor | 0.976 | 0.942 |
| Immediacy Index | 0.890 | 0.792 |
| Cited Half-life | 0.900 | 0.929 |
| | | |
| ***New Indicators*** | | |
| H-index (Scopus) | 0.995 | 0.984 |
| SJR | 0.990 | 0.947 |
| PageRank ($d = 0.85$) | 0.828 | 0.841 |
| PageRank ($d = 0.15$) | 0.886 | 0.870 |
| | | |
| ***Size Indicators*** | | |
| Total Cited ISI | 0.999 | 0.989 |
| Total Cited Scopus | 0.989 | 0.969 |
| Total Citing ISI | 0.824 | 0.914 |
| Total Citing Scopus | 0.970 | 0.903 |
| Self-citations ISI | 0.989 | 0.926 |
| Total Docs. ISI | 0.967 | 0.937 |
| Total Docs. Scopus | 0.964 | 0.922 |
| | | |
| ***Network indicators*** | | |
| Indegree (cited) | 0.998 | 0.988 |
| Outdegree (citing) | 0.963 | 0.941 |
| Betweenness (cited) | 0.999 | 0.960 |
| Betweenness (citing) | 0.905 | 0.830 |
| Closeness (cited) | 0.794 | 0.941 |
| Closeness (citing) | 0.611 | 0.866 |

**Table 7**: Auto-correlations between the corresponding indicator values in 2006 and 2007.

All correlations are significant at the 0.01 level; $N = 5,861$ except in the case of Cited Half-life (because of the many missing values in this case: $N = 4,701$). The main observation is that the new journal indicators (SJR and H-index) are more stable ($r > 0.99$) than the traditional ISI-indicators. As expected, all indicators in the citing dimension are less stable than in the cited one because citing as action introduces change into the database (Leydesdorff, 1993). Closeness centrality is less stable than the other network indicators ($0.61 < r < 0.79$).

**Conclusions**

Unlike the Impact Factor, the SJR indicator and the H-index are non-parametric, that is, they are not based on normalization to the arithmetic mean of the values in the document sets. In the case of citation data this is a clear advantage, because the data is heavily skewed (Leydesdorff, 2008). However, the H-index has the well known problem that it leads to counter-intuitive results because of the attempt to bring the size component and the impact component under a single denominator. For example, an author who has published three articles with 100 citations each and otherwise a number of articles cited fewer than three times is equally ranked at $h = 3$ with an author who has only marginally more than three citations for each of the top three articles and otherwise the same distribution. The advantage of the H-index is that it provides a single number which is easy to retrieve and remember.

The data contains two dimensions, size and impact, which—quoting Garfield—together shape influence. Any inference which jumps too easily from one dimension to another may lead to misunderstandings. As Garfield emphasized, one should not consider the Impact Factor as an indicator of the quality of a journal, but as an indicator of the average quality of articles in that journal. A comparison with the size dimension makes this point clear: the average size of an article does not inform us about the size of the journal. However, this warning against an "ecological fallacy"—that is failing to note the difference between making inferences about the aggregate of trees or of the woods—has hitherto been little heeded in studies using the Impact Factor for research evaluation (Robertson, 1950; Kreft & De Leeuw, 1988; Moed, 2005). Journals, however, are *not* homogenous entities (Bradford, 1934; Garfield, 1971; Bensman, 2007).

I was surprised to find that PageRank was to such an extent *not* an impact indicator because it was developed after Pinski & Narin's (1976) "Influence Weights," which were designed as an alternative to Impact Factors. In another context, Hindman *et al.* (2003) found supporting evidence for these size-effects in the results of Google's algorithm. However, the SJR, albeit derived conceptually from the PageRank concept, seems to offer a good alternative to the Impact Factor.

The Cited Half-life provides a separate dimension for the evaluation. I showed elsewhere (Leydesdorff, 2008, at p. 280) that this indicator enables us to distinguish different expected citation behaviors among sets based on different document types (articles, reviews, and letters) independent of the differences in citation behavior among disciplines. Based on a suggestion of Sombatsompop *et al.* (2004), Rousseau (2005) proposed to elaborate on the Cited Half-life for developing median and percentile impact factors as another set of new indicators.

Indicators along the impact axis, including the newly introduced Audience Factor (Zitt & Small, 2008), are highly correlated. In an ideal world, one might perhaps like to see the SJR applied to the JCR data (Bergstrom, 2007). One advantage of the SJR is that this indicator is available as open access—although the Impact Factor values are so much circulated in practice (e.g., at http://abhayjere.com/impactfactor.aspx ) that one can hardly consider this a serious drawback in using them. An advantage of the Impact Factor and the H-index, on the other hand, is the ease with which they can be understood. As I have shown, however, the H-index may oversimplify the complexities involved because it tries to capture both orthogonal dimensions (size and impact) in a single indicator.

**Acknowledgements**

I am grateful to Han Park and Jeong-Soo Seo for providing the results of the PageRank analysis using NetMiner, and to Michel Zitt and Henry Small for kindly providing the data of the Audience Factor for 2006. An anonymous referee provided suggestions for improvements of the paper.